\documentstyle[11pt,epsf]{article}

\def\ni{\noindent}
\begin{document}
\vskip-4cm
\phantom{1234}\hfill SUNY-NTG-95-52 \\
\phantom{1234}\hfill December, 1995

\vspace*{0.3in}
\begin{center}
  {\bf NEUTRINO SCATTERING IN STRANGENESS-RICH STELLAR MATTER } \\
  \bigskip
  \bigskip
  SANJAY REDDY AND MADAPPA PRAKASH \\
  \bigskip
        {\footnotesize Physics Department \\
        State University of New York at Stony Brook\\
        Stony Brook, NY 11794-3800, USA }
  \bigskip
\end{center}
\smallskip
\centerline{ABSTRACT}
\begin{quote}
{\small We calculate neutrino cross sections from neutral current reactions
in dense matter containing hyperons.    We show that $\Sigma^-$ hyperons give
significant contributions.  To lowest order, the contributions from the
neutral $\Lambda$ and $\Sigma^0$, which have zero hyper-charge, vanish.
However, their presence furnishes baryon number which decreases the relative
concentrations of nucleons.  This leads to significant reductions in the cross
sections. Due to the uncertainty  in strong interactions at high density,  the
neutrino opacity may vary by a  factor of about 2 depending on the behavior of
the effective masses.
}
\end{quote}
\ni {\em Subject headings:} dense matter -- stars: neutron -- stars: opacities
-- stars: neutrinos

\vspace*{0.3in}

\section{Introduction}

The general nature of the neutrino signature expected from a newly formed
neutron star (hereafter referred to as a protoneutron star) has been
theoretically predicted~(Burrows \& Lattimer 1986) and  confirmed by the
observations~(Bionta et al. 1987; Hirata et al 1987)  from supernova SN1987A.
Although neutrinos interact weakly with matter, the high baryon densities and
neutrino energies achieved after the gravitational collapse of a massive star
($\geq 8$ solar masses) cause the neutrinos to become trapped on the dynamical
timescales of collapse~(Sato 1975; Mazurek 1975).  Trapped neutrinos at the
star's core have  Fermi energies $E_\nu \sim 200-300$ MeV and are primarily of
the $\nu_e$ type.  They escape after diffusing through the star exchanging
energy with the ambient matter, which has an entropy per baryon of order unity
in units of Boltzmann's constant.  Eventually they emerge from the star with an
average energy $\sim 10-20$ MeV and in nearly equal abundance of all three
flavors, both particle and anti-particle.

Neutrino interactions in dense matter have been investigated by various
authors (Tubbs \& Schramm 1975; Sawyer 1975,89,95; Lamb \& Pethick 1976;
Lamb 1978; Sawyer \& Soni 1979; Iwamoto \& Pethick 1982; Iwamoto 1982;
Goodwin \& Pethick 1982; Burrows \& Mazurek 1982; Bruenn 1985;
van den Horn \& Cooperstein 1986; Cooperstein 1988; Burrows 1988; Horowitz
\& Wehrberger 1991a,b;1992; Reddy \& Prakash 1995).
The charged current absorption and neutral current scattering reactions are
both important sources of opacity. The neutral current scattering involves all
flavors of neutrinos scattering on nucleons and leptons. Scattering from
electrons is important for energy and momentum transfer (Tubbs \& Schramm,
1975).  The  influence of interactions for neutrino-electron scattering is also
important (Horowitz 1992)  and increases the mean free path by 50-60\% for
electron type neutrinos.  However, for lepton number transport, nucleon
scattering and absorption are the dominant processes.

Surprisingly little attention has been paid to the effects of composition and
of strong interactions of the ambient matter on neutrino opacities.  The effect
of interactions was investigated for non-degenerate nuclear matter by Sawyer
(1975; 1989) and for degenerate pure neutron matter by Iwamoto
\& Pethick (1982).
Treating nucleons in the non-relativistic limit, these calculations predict an
increase in the mean free path by a factor of  $\sim 2-3$, for (2-4) times the
nuclear density.  More recently, relativistic calculations based  on  effective
Lagrangian models for dense neutron star  matter have been performed by
Horowitz \& Wehrberger (1991a,b;1992).  Here, the differential cross sections
for matter  containing nucleons and  electrons were  calculated using linear
response theory.  A reduction of 30-50\% over the case of non-interacting
nucleons was reported in these calculations.   The influence of interactions
has been investigated in protoneutron star calculations only by a simple
scaling of the non-interacting results (Burrows 1990, Keil 1994).
Furthermore, there have been no calculations performed including the
multi-component nature of the system.  We note that Keil \& Janka (1995) have
recently carried out cooling simulations including hyperons in the equation of
state (EOS), but they ignored opacity modifications.  We view it as essential
that opacities be consistent with the composition, which has not been a feature
of protoneutron star models to date.

Although the composition and EOS  of the hot
protoneutron star matter are not yet known with certainty, QCD based effective
Lagrangians have opened up intriguing possibilities (Kaplan \& Nelson 1986;
Glendenning 1986,1992;  Glendenning \& Moszkowski 1991; Kapusta \& Olive
1990; Ellis, Knorren \&  Prakash 1995; Knorren, Prakash \& Ellis 1995,
Prakash, Cooke \& Lattimer 1995). Among these is the possible existence  of
matter with strangeness to baryon ratio of order unity.  Strangeness may be
present either in the form of fermions, notably the $\Lambda$ and $\Sigma^-$
hyperons, or, in the form of a Bose condensate, such as a $K^-$- meson
condensate, or, in the form of $s$ quarks.   In the absence  of trapped
neutrinos,  strange particles are expected to appear around $2-4$  times the
nuclear matter density of $n_0=0.16~{\rm fm}^{-3}$.   Neutrino-trapping causes
the  strange particles to appear at somewhat  higher densities, since the
relevant chemical potential $\mu = \mu_e -\mu_{\nu_e}$ in matter with high
lepton content is much smaller than in the untrapped case (Ellis, Knorren \&
Prakash 1995; Knorren, Prakash \& Ellis 1995).

A new feature that we consider here is the role of strangeness.  To date, only
neutrino opacities for strange quark matter have been calculated (Iwamoto,
1982).  Here, we study neutrino mean free paths in matter containing
strangeness in the form of hyperons.  Specifically, we calculate neutrino
opacities from neutral current reactions  in matter containing hyperons  and
which are faithful to the EOS. In a first effort, this will be achieved using a
mean field theoretical description which includes hyperonic degrees of freedom.
This approach has several merits.  For example,  aspects of relativity, which
may become important at high density, are  naturally incorporated.
Modifications of the opacity due to correlations (RPA) are also possible in
such an approach.   Further, comparisons with alternative potential model
approaches (Iwamoto \& Pethick 1982; Sawyer 1989) are straightforward.
Neutrino opacities in matter containing other forms of strangeness and from
charged current reactions (Prakash et. al. 1992) will be considered in
a separate work.

In \S 2, neutrino interactions with strange baryons are discussed. In \S 3,
the composition of beta-equilibrated matter with strange baryons is
determined based on a field theoretic description.  \S 4 contains our
results along with discussion.  Conclusions are given in \S 5.

\section{ Neutrino Interactions with Strange Baryons }

Neutrino interactions with matter proceed via charged and neutral
current reactions. The neutral current processes contribute to
elastic scattering, and charged current reactions result in neutrino
absorption.  The formalism to calculate neutral current scattering rates in
dense matter is summarized below.
The interaction Lagrangian for neutrino
scattering reactions is given by the Wienberg-Salam theory:
\begin{eqnarray}
{\cal L}_{int}^{nc} &=& ({G_F}/{2\sqrt{2}}) ~~l_\mu
j_z^\mu\, \qquad {\rm ~for} \qquad \nu + B \rightarrow  \nu + B \,,
\end{eqnarray}
where $G_F\simeq 1.436\times 10^{-49}~{\rm erg~cm}^{-3}$ is the weak coupling
constant.  The neutrino and target particle weak neutral currents appearing
above are:
\begin{eqnarray}
l_\mu^\nu &=& {\overline \psi}_\nu \gamma_\mu
\left( 1 - \gamma_5 \right) \psi_\nu \nonumber\\
j_z^\mu &=& {\overline \psi}_i \gamma^\mu
\left( C_{Vi} - C_{Ai} \gamma_5 \right) \psi_i \,,
\end{eqnarray}
where $i=n,p,\Lambda,\Sigma^-,\Sigma^+,\Sigma^0,\Xi^-,\cdots$ and
$e^-,\mu^-$.  The
neutral current process couples neutrinos of all types ($e,\mu$ and $\tau$) to
the  weak neutral hadronic current, $j_z^\mu$.  The vector and axial vector
coupling constants, $C_{Vi}$ and $C_{Ai}$,  are listed in Table 1. Numerical
values of the parameters that best fit data on  charged current semi-leptonic
decays of hyperons are (Gaillard \& Sauvage 1984):  D=0.756 , F=0.477,
$\sin^2\theta_W$=0.23 and $\sin\theta_c = 0.231$.  Tree level coupling of
neutrinos to the neutral particles $\Lambda$ and $\Sigma^0$ vanish, since the
Z boson couples to the hyper-charge, which is zero for both the  $\Lambda$
and $\Sigma^0$.  Neutrino scattering off leptons in the same family involves
charged current couplings as well, and one has to sum over both the
contributing diagrams.  At tree level, however, one can express the total
coupling by means of a Fierz transformation; this is accounted for in Table 1.

\begin{table}
\begin{center}
\vskip 10pt
\centerline{TABLE 1}
\vskip 5pt
\centerline{NEUTRAL CURRENT VECTOR AND AXIAL COUPLINGS}
\vskip 5pt
\begin{tabular}{lcc}
\hline \hline
{Reaction } & $C_V$  & $C_A $  \\
\hline
$ \nu_e + e \rightarrow \nu_e + e$ & $ 1+4\sin^2\theta_W=1.92$ & $1$\\
$\nu_e + \mu \rightarrow \nu_e + \mu$ & $ -1+4\sin^2\theta_W=-0.08$ & $-1$\\
$\nu_i + n \rightarrow \nu_i + n$ & $-1$ & $-D-F=-1.23$ \\
$\nu_i + p \rightarrow \nu_i+ p$ & $ 1-4\sin^2\theta_W=0.08$ & $D+F=1.23$
  \\
$\nu_i + \Lambda\rightarrow \nu_i +\Lambda$ & 0& 0\\
$\nu_i + \Sigma^-\rightarrow \nu_i +\Sigma^-$ &$ -2+4\sin^2\theta_W=-1.08$ &
 $-2F=-0.95$  \\
$\nu_i + \Sigma^+\rightarrow \nu_i +\Sigma^+$&$ 2-4\sin^2\theta_W=1.08$ & 2F
=0.95 \\
$\nu_i + \Sigma^0\rightarrow \nu_i +\Sigma^0$ & $0$ & $0$\\
$\nu_i + \Xi^-\rightarrow \nu_i +\Xi^-$ & $ -1+4\sin^2\theta_W=-0.08 $ & D
=0.756 \\
$\nu_i + \Xi^0\rightarrow \nu_i +\Xi^0$ & 1 & $-D+F=-0.28$ \\
\hline \hline
\end{tabular}
\begin{quote}
{\footnotesize NOTE.-- Coupling constants derived assuming SU(3) symmetry for
the hadrons.  Numerical values are quoted using  D=0.756 , F=0.477,
$\sin^2\theta_W$=0.23 and $\sin\theta_c = 0.231$ (Gaillard \& Sauvage 1984).
At tree level, the $\Lambda$ and $\Sigma^0$, which have  zero weak-hypercharge,
do not couple to the neutrinos.}
\end{quote}

\end{center}
\end{table}

Given the general structure of the neutrino coupling to matter,
the differential cross section for elastic scattering for incoming neutrino
energy $E_{\nu}$ and outgoing neutrino energy $E_{\nu}^{'}$ is given by
\begin{eqnarray}
\frac{1}{V}\frac{d^3\sigma}{d\Omega^{'2} dE_{\nu}^{'}}& =& -\frac{G^2}{128
\pi^2}\frac{E_{\nu}^{'}}{E_{\nu}}~{\rm Im}~(L_{\alpha\beta}\Pi^{\alpha\beta})
\,,
\label{dcross}
\end{eqnarray}
where the neutrino tensor $L_{\alpha\beta}$ and the target particle
polarization $\Pi^{\alpha\beta}$ are
\begin{eqnarray}
L_{\alpha\beta} &=& 8[2k_{\alpha}k_{\beta}+(k\cdot q)g_{\alpha\beta}
-(k_{\alpha}q_{\beta}+q_{\alpha}k_{\beta})\mp i\epsilon_{\alpha\beta\mu\nu}
k^{\mu}q^{\nu}]\\
\Pi_{\alpha\beta}^i & = & -i \int \frac{d^4p}{(2\pi)^4} \,
{\rm Tr}~[G^i(p)J_{\alpha} G^i(p+q)J_{\beta}]\,.
\end{eqnarray}
Above, $k_{\mu}$ is the incoming neutrino four momentum, and $q_{\mu}$ is  the
four momentum  transfer.    The Greens' functions $G^i(p)$ (the index $i$
labels particle species)  depend on the Fermi momentum $k_{Fi}$ of target
particles.  In the Hartree approximation, the  propagators are obtained by
replacing $M_i$ and $k_{Fi}$  in the free particle propagators by $M^*_i$ and
$k^*_{Fi}$ (see below), respectively.  The current operator $J_{\mu}$ is
$\gamma_{\mu}$ for the vector current and $\gamma_{\mu}\gamma_5$ for the axial
current. Given the V--A structure of the particle currents, we have
\begin{eqnarray}
\Pi_{\alpha\beta}^i
&=&C_{Vi}^2\Pi_{\alpha\beta}^{V~i}+C_{Ai}^2\Pi_{\alpha\beta}^{A~i}
-2C_{Vi}C_{Ai}\Pi_{\alpha\beta}^{VA~i} \,.
\end{eqnarray}
For the vector polarization,
$\{J_\alpha,J_\beta\}::\{\gamma_\alpha,\gamma_\beta\}$,
for the axial
polarization,  $\{J_\alpha,J_\beta\} ::
\{\gamma_\alpha\gamma_5,\gamma_\beta\gamma_5\}$ and for the mixed
part,  $\{J_\alpha,J_\beta\} ::\{\gamma_\alpha\gamma_5,
\gamma_\beta\}$.  Further, the polarizations contain two parts:
the density dependent part that describes particle-hole excitations and
the Feynman part that describes particle-antiparticle excitations.  For elastic
scattering,  with $q_{\mu}^2<0$,   the contribution of the  Feynman parts
vanish.   Using vector current conservation and translational invariance,
$\Pi_{\alpha\beta}^V$ may be written in terms of two independent components.
In a frame where $q_{\mu}=(q_0,|q|,0,0)$, we have
\begin{eqnarray*}
\Pi_T = \Pi^V_{22} \qquad {\rm and} \qquad
\Pi_L = -\frac{q_{\mu}^2}{|q|^2}\Pi^V_{00} .\\
\end{eqnarray*}
The axial current-current correlation function can be written as a vector
piece plus a correction term:
\begin{eqnarray}
\Pi_{\mu \nu}^A=\Pi^V_{\mu \nu}+g_{\mu \nu}\Pi^A .\
\end{eqnarray}
The mixed, axial current-vector current correlation function is
\begin{eqnarray}
\Pi_{\mu \nu}^{VA}= i\epsilon_{\mu, \nu,\alpha,0}q^{\alpha}\Pi^{VA}. \
\end{eqnarray}
The above mean field or Hartree polarizations, which characterize the
response of the medium to the neutrino, have been explicitly evaluated in
previous work (Horowitz \& Wehrberger 1991).  In terms of these polarizations,
the  differential cross section is
\begin{eqnarray}
\frac{1}{V}\frac{d^3\sigma}{d\Omega^{'2} dE_{\nu}^{'}}& =& -\frac{G^2}{16
\pi^3}~ \frac{E_{\nu}^{'}}{E_{\nu}} q_{\mu}^2 ~[AR_1+R_2+BR_3]\,,
\label{dcross2}
\end{eqnarray}
with
\begin{eqnarray}
A=\frac{2k_0(k_0-q_0)+q_{\mu}^2/2}{|q|^2}~~
;~~B=2k_0-q_0.\
\end{eqnarray}
The polarizations may be combined into three uncorrelated response functions,
$R_1,R_2$ and $R_3$, by summing over the contributions from each particle
species $i$:
\begin{eqnarray}
R_1&=&\sum_{i}~[C_{Vi}^2+C_{Ai}^2][Im(\Pi_T^i) + Im(\Pi_L^i)],\\
{\label{r1}}
R_2&=&\sum_{i}~ C_{Vi}^2 [Im(\Pi_T^i)] + C_{Ai}^2
[Im(\Pi_T^i) - Im(\Pi_A^i)],\\
{\label{r2}}
R_3&=&\pm \sum_{i}~ 2 C_{Ai}C_{Ai}~Im(\Pi_{VA}^i)\,.
{\label{r3}}
\end{eqnarray}
These functions depend upon the the individual
$k_{Fi}$ (or the concentration)  and the corresponding effective masses
$M_i^*$, for which a many-body  description of the multi-component system is
required.

\section{ Composition of Matter with Strange Baryons }

To explore the influence of the presence of
hyperons in dense matter, we employ a relativistic
field theoretical model in which the interactions between baryons are mediated
by the exchange of $\sigma,\omega$ and $\rho$ mesons.  The full Lagrangian
density is given by~ (Serot \& Walecka 1986),
\begin{eqnarray*}
L= &\sum_{B}& \overline{B}(-i\gamma^{\mu}\partial_{\mu}-g_{\omega B}
\gamma^{\mu}
-g_{\rho B}\gamma^{\mu}{\bf{b}}_{\mu}\cdot{\bf t}-M_B+g_{\sigma B}\sigma)B \\
&-& \frac{1}{4}W_{\mu\nu}W^{\mu\nu}+\frac{1}{2}m_{\omega}^2\omega_{\mu}\omega^
{\mu} - \frac{1}{4}{\bf B_{\mu\nu}}{\bf
B^{\mu\nu}}+\frac{1}{2}m_{\rho}^2 \rho_{\mu}\rho^{\mu}\\
&+& \frac{1}{2}\partial_{\mu}\sigma\partial^{\mu}\sigma +\frac{1}{2}
m_{\sigma}^2\sigma^2-U(\sigma)\\
&+& \sum_{l}\overline{l}(-i\gamma^{\mu}\partial_{\mu}-m_l)l \,.
\end{eqnarray*}
Here, $B$ are the Dirac spinors for baryons and $\bf t$ is the isospin
operator. The sums include baryons, $B=n,p,\Lambda,\Sigma$ and $\Xi$, and
leptons, $l = e^-$ and $\mu^-$. The field strength tensors for the $\omega$
and
$\rho$ mesons are $W_{\mu\nu} = \partial_\mu\omega_\nu-\partial_\nu\omega_\mu$
and ${\bf B}_{\mu\nu} =  \partial_\mu{\bf b}_\nu-\partial_\nu{\bf b}_\mu$ ,
respectively.  The potential $U(\sigma)$ represents the self-interactions of
the scalar field and is taken to be of the form
\begin{eqnarray}
U(\sigma) =  \frac{1}{3}bM_n(g_{\sigma N}\sigma)^3 + \frac{1}{4}c(g_{\sigma
N}\sigma)^4\,.
\end{eqnarray}
Electrons and muons are included in the model as non-interacting particles,
since their interactions give small contributions compared to those of
their free Fermi gas parts.

In the mean field approximation, the baryon source currents  and meson fields
are replaced by their ground state expectation values $\sigma_0, \omega_0$ and
$b_0$.  The resulting set of non-linear equations are solved for the meson
fields  and the particle fractions under the constraints of  charge neutrality
and $\beta$- equilibrium.  At zero temperature, the particle fractions $n_i
\propto k_{Fi}^3$. The general beta equilibrium condition
\begin{eqnarray}
\mu_i = b_i\mu_n - q_i\mu_l\,,
\end{eqnarray}
where $b_i$ is the baryon number of particle $i$ and $q_i$ is its
charge, determines whether a particular baryon will be present at a given
density. This will be the case if the lowest lying energy state of that baryon
in matter, which is implicitly density dependent through the values of the
meson fields, is less than its chemical potential as dictated by
$\beta$-equilibrium. The value of $k_{FB}$ is thus determined by the
requirement that
\begin{eqnarray}
\mu_B = e_B(k_{FB}) = g_{\omega B}\omega_0 + g_{\rho B} t_{3B}b_0
+ {\sqrt {k_{FB}^2 + M_B^{*2}}} \, ,
\label{thresh}
\end{eqnarray}
where the Dirac effective mass $M_B^* = M_B - g_{\sigma B} \sigma_0$.

In the nucleon sector, the constants $ g_{\sigma N}, g_{\omega N},
g_{\rho N}, b$ and $c$ are determined by reproducing the nuclear matter
equilibrium density $n_0=0.16~{\rm fm}^{-3}$, and the binding energy per
nucleon,  the symmetry energy coefficient, the compression modulus, and the
nucleon Dirac effective mass, $M^*$, at $n_0$.  Numerical values of the
coupling
constants so chosen in  different models are shown in Table 2. Models from
Glendenning and Moszkowski (1991) are termed GM and models from Horowitz and
Serot (1981) are termed HS. The different values shown reflect the
prevalent uncertainty in the  nuclear matter compression modulus
and the effective mass $M^*$.

\begin{table}
\vskip 5pt
\centerline {TABLE 2}
\vskip 5pt
\centerline{NUCLEON-MESON COUPLING CONSTANTS  }
\begin{center}
\begin{tabular}{ccccccc}
\hline \hline
Model & $\frac{g_{\sigma}}{m_{\sigma}}$
&$\frac{g_{\omega}}{m_{\omega}}$&$\frac{g_{\rho}}{m_{\rho}}$& b & c&
$\frac{M^*}{M}$ \\
 &(fm)  &(fm)  &(fm)  &  &   & \\
\hline
GM1 & 3.434 &2.674  &2.100  &0.00295  &-0.00107   &0.70 \\
GM2 &3.025  &2.195 &2.189 &0.00348 &0.01328 &0.78 \\
GM3 &3.151  &2.195 &2.189 &0.00866 &-0.00242 &0.78 \\
HS1,HS2,HS3 &3.974  &3.477 &2.069 &0.0 &0.0 &0.54 \\
\hline \hline
\end{tabular}
\begin{quote}
{\footnotesize NOTE.-- Constants from
Glendenning and Moszkowski (1991) are termed GM and those from
Horowitz and Serot (1981) are termed HS.  HS models do
not have scalar self-couplings which leads to effective masses which are
significantly smaller than those of GM models.}
\end{quote}
\end{center}
\end{table}

In the GM models, the hyperon coupling constants are determined by
reproducing the binding energy of the $\Lambda$ hyperon in nuclear matter.
Parameterizing the hyperon-meson couplings in terms of nucleon-meson couplings
through
\begin{eqnarray}
x_{\sigma H}=g_{\sigma H}/g_{\sigma N},~~~
x_{\omega H}=g_{\omega H}/g_{\omega N}
,~~~x_{\rho H}=g_{\rho H}/g_{\rho N} \,,
\end{eqnarray}
the $\Lambda$ binding energy at nuclear density is given by $(B/A)_\Lambda =
-28 = x_{\omega \Lambda} g_{\omega N} \omega_0 - x_{\sigma \Lambda} g_{\sigma
N} \sigma_0$, in units of MeV.  Thus, a particular choice of  $x_{\sigma
\Lambda}$ determines $x_{\omega \Lambda}$ uniquely.   To keep the number of
parameters small, the coupling constant
ratios for all the different hyperons are assumed to be the same. That is
\begin{eqnarray}
x_\sigma = x_{\sigma\Lambda} = x_{\sigma\Sigma} = x_{\sigma\Xi} \,,
\end{eqnarray}
and similarly for the $\omega$ and $\rho$ mesons. Further,
$x_\rho$ is set equal to $x_\sigma$.

In a recent analysis of $\Sigma^-$ atoms, Mare\u{s} et al. (1995) obtain
reasonable fits with $x_{\omega\Sigma}= 2/3$  and 1, and $x_\sigma\Sigma =
0.54$ and 0.77 based on a mean field description of nuclear matter using the
nucleon couplings of Horowitz \& Serot (1981).  Lacking further inputs
from data,  Mare\u{s} et al. assume that the couplings of the
$\Sigma$ and $\Xi$ are equal to those of the $\Lambda$ hyperon. This set of
couplings are termed HS1 in Table 3. Following Knorren, Prakash \& Ellis
(1995), we have relaxed the above assumption about the couplings of the
$\Sigma$ and $\Xi$  in parameter sets HS2 and HS3 to explore the sensitivity of
the thresholds to small changes in the couplings.

\begin{table}
\vskip 5pt
\centerline {TABLE 3}
\vskip 5pt
\centerline{ RATIOS OF HYPERON-MESON TO NUCLEON-MESON COUPLING CONSTANTS }
\begin{center}
\begin{tabular}{cccccccccc}
\hline \hline
Model & $x_{\sigma\Lambda }$&$x_{\sigma\Sigma }$&$x_{\sigma\Xi }$
& $x_{\omega\Lambda }$&$x_{\omega\Sigma }$&$x_{\omega\Xi }$
& $x_{\rho\Lambda }$&$x_{\rho\Sigma }$&$x_{\rho\Xi }$\\
\hline
GM1  &0.600  &0.600 &0.600  &0.653  &0.653  &0.653  &0.600   &0.600 &0.600 \\

GM2,GM3  &0.600  &0.600 &0.600  &0.659  &0.659  &0.659  &0.600 &0.600 &0.600 \\

HS1  &0.600  &0.540 &0.600  &0.650  &0.670  &0.650  &0.600 &0.670 &0.600 \\

HS2  &0.600  &0.770 &0.600  &0.650  &1.00  &0.650  &0.600 &0.670 &0.600 \\

HS3  &0.600  &0.770 &0.770  &0.650  &1.00  &1.00  &0.600 &0.670 &0.670 \\
\hline \hline
\end{tabular}
\begin{quote}
{\footnotesize NOTE.-- $x_{iH}=g_{iH}/g_{iN}$, where $i=\sigma,\omega$ or
$\rho$ and $H$ is a hyperon.}
\end{quote}
\end{center}
\end{table}

In Figure~1, we show the relative fractions of the baryons and leptons
in charge neutral and $\beta-$equilibrated matter.  For the GM models, the
$\Sigma^-$ hyperon appears at a density lower than the $\Lambda$ hyperon.  This
is because the somewhat higher mass of the $\Sigma^-$ is compensated by the
presence of the $e^-$ chemical potential in the equilibrium condition of the
$\Sigma^-$. More massive and more positively charged particles appear at
higher densities. With the appearance of the negatively charged $\Sigma^-$,
which competes with leptons in maintaining charge neutrality, the lepton
concentrations begin to fall.  The important point is that, with increasing
density, the system contains many baryon species with nearly equal
concentrations.

\begin{figure}
\begin{center}
\leavevmode
\epsfxsize=3.0in
\epsfysize=3.5in
\epsffile{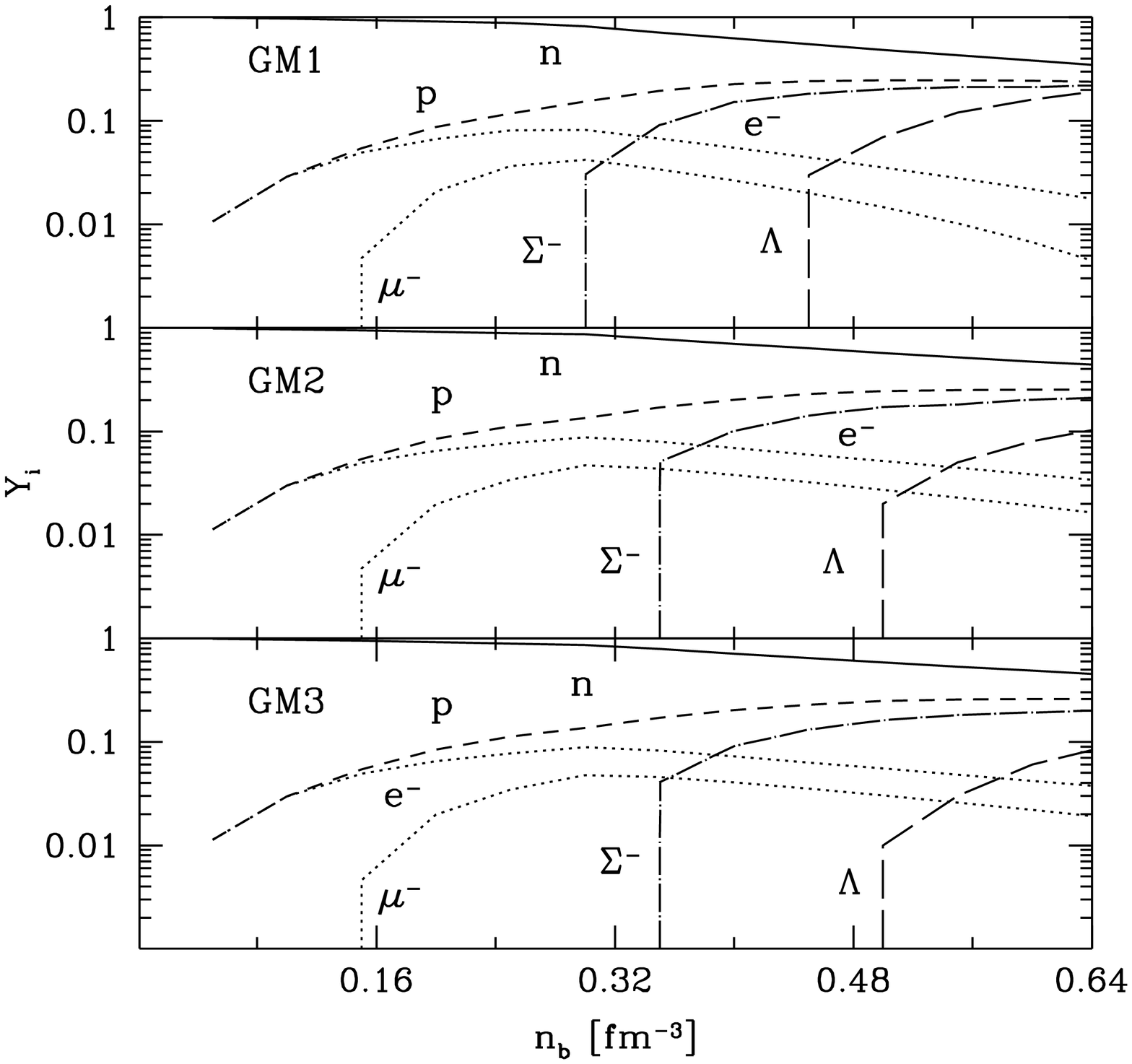}
\epsfxsize=3.0in
\epsfysize=3.5in
\epsffile{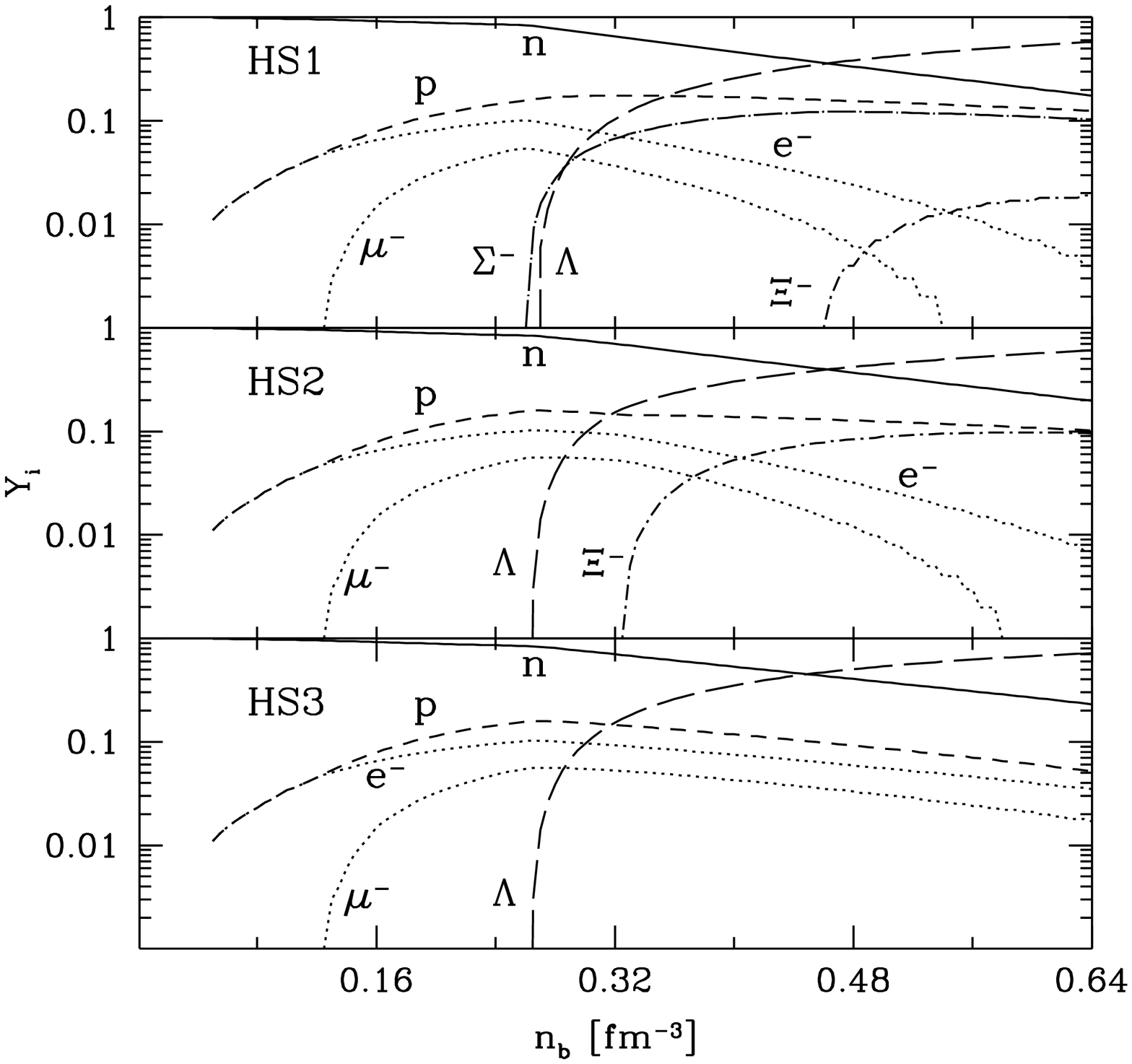}
\end{center}
\caption[]{\footnotesize
Particle fractions, $Y_i=n_i/n_b$, for the models shown in Table 3.  }
{\label{pfracs}}
\end{figure}

The relative concentrations of the baryons in the HS1 model are qualitatively
similar to those of the GM models, although substantial
quantitative differences exist.
Note, however, that relaxing the assumption about the $\Sigma$ and $\Xi$
couplings has large effects on the appearance of negatively charged particles,
as seen from the results of models HS2 and HS3.   Increasing the
coupling constants of a hyperon species delays its appearance to a higher
density.  This is because  the threshold equation, Eq.~(\ref{thresh}), receives
contributions from the $\sigma, \omega$ and $\rho$ fields, the net result being
positive due to that of the $\omega$. If all the couplings are scaled up, the
positive contribution becomes larger, and hence the appearance of the particle
is delayed to a higher density. The $\Sigma$ couplings of set HS2 are larger
than those of set  HS1, so the $\Sigma^-$ no longer appears, thus allowing the
chemical potential $\mu$ to continue rising. This allows the $\Xi^-$ to appear
at $n/n_0 \cong 2.2$, essentially substituting for the $\Sigma^-$.
Were we to reduce the $\Xi$ couplings on the grounds that this hyperon contains
two strange quarks, the $\Xi^-$ would appear at an even lower density.  In
model HS3, both the $\Sigma$ and $\Xi$ couplings are increased. Neither of them
now appear, leaving the $\Lambda$ as the only strange particle in matter.
Clearly, the thresholds for the strange particles are sensitive to the coupling
constants, which are presently poorly constrained by either theory or
experiment.  Notwithstanding these caveats,  it is clear that one or the other
hyperon species is likely to exist in dense matter.

As we will see later, the effective masses of the baryons play an important
role in determining the differential cross sections.  Figure~2
shows the  density dependence of the  nucleon effective mass, which  shows
significant differences between models with  scalar self-interactions (GM) and
those without (HS).  Scalar self-interactions, which lead to low $\sigma N$
couplings in GM models, result in $M^*$s which are larger than those of the HS
models, in which $M^*$ falls rapidly, due to a  relatively large $\sigma N$
coupling.   The presence of hyperons further hastens this fall off with
density, which is evident from the density  dependence of $M^*$ in matter with
nucleons and leptons only (see the curve labeled HS*).   The behavior of the
effective masses tending to zero  at a finite baryon density is generic to
models with hyperons, as shown by Knorren, Prakash \& Ellis (1995).  Whether
this feature may be interpreted as  strangeness-induced chiral restoration
depends on whether the effective mass can be viewed as an order parameter.  In
the context of neutrino interactions, which is our main interest here,  the
results of Figure~2 may be taken to encompass the current
uncertainty in the knowledge of strong interactions.

\begin{figure}
\begin{center}
\leavevmode
\epsfxsize=3.0in
\epsfysize=3.0in
\epsffile{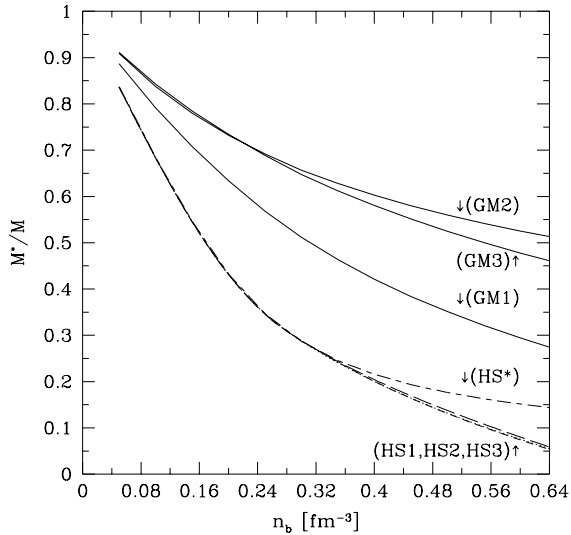}
\end{center}
\caption[]{\footnotesize
Nucleon effective mass $M^*$ for the models discussed in the text.  The
plot labelled HS* is for matter containing nucleons and leptons only, with
couplings of model HS.}
{\label{efmass}}
\end{figure}

\section { Results and Discussion }

Since the neutrino coupling is species specific (see Table~1), the response of
a multi-component system will depend on the relative abundance of the various
particles.   Pauli blocking, which plays an important role at
zero temperature, restricts the neutrinos to couple with particles lying  close
to their respective Fermi surfaces.   The Fermi momenta and the effective
masses of the various particles  in different models are given in Table~4.
Results here are for $n_b=0.4~{\rm fm}^{-3}$.

\begin{table}
\vskip 5pt
\centerline {TABLE 4}
\vskip 5pt
\centerline{FERMI MOMENTA AND EFFECTIVE MASSES}
\vskip 5pt
\begin{center}
\begin{tabular}{lcccccccccc}
\hline \hline
Model & $n_b$ & $k_{Fn}$ & $k_{Fp}$ & $k_{F\Sigma^-}$ & $k_{F\Lambda}$ &
$k_{Fe^-}$ & $k_{F\mu^-}$ & $M_n^*$ & $M_{\Sigma}^*$ & $M_{\Lambda}^*$ \\

&$fm^{-3}$ & MeV & MeV & MeV & MeV & MeV & MeV & MeV & MeV & MeV \\
\hline

GM1 & 0.4 & 385.2 & 274.2 & 236.3 & 0.0 & 171.1 & 134.3 & 395.4 & 867.4 &
790.5 \\

GM3 & 0.4 & 401.0 & 263.9 & 201.1 & 0.0 & 187.2 & 154.3 & 545.4 & 957.4 &
880.5 \\

HS1 & 0.4 &348.4 &249.0 &214.3 & 286.5 & 158.2 & 117.7 & 188.1 & 788.1 & 666.1
 \\

HS3 & 0.4 & 364.4 & 220.4 & 0.0 & 317.2 & 189.6& 157.4 & 191.8 & 618.4 & 668.3
\\
\hline \hline
\end{tabular}
\begin{quote}
{\footnotesize NOTE.-- Results are for $n_b=0.4~{\rm fm}^{-3}$.}
\end{quote}
\end{center}
\end{table}

Figure~3 shows the response function $R_1$ at a density  $n_b=0.4~{\rm
fm}^{-3}$.   Models HS* and GM3* in the left panels are for matter with
nucleons and  leptons, while models HS3 and GM3 in the right panels include
hyperons.   For a given three-momentum transfer $|q|$,  each particle species
provides support  to the response functions in the region
$0\le\omega\le\omega_{max}$,  where $\omega_{max}$ for particle-hole
excitations is determined by  energy-momentum conservation:  \begin{eqnarray*}
\omega_{max} =\sqrt{E_{Fi}^{*2}+q^2+2k_{Fi}|q|} - E_{Fi}^* \,, \end{eqnarray*}
where $E_{Fi}^* = {\sqrt {k_{Fi}^2 + M_i^{*2}}}$.  For $M_i^* \gg k_{Fi}$, this
reduces to the non-relativistic condition $\omega_{max} = ({k_{Fi}}/{M_i^*})
|q|$.   The response functions for electron and muon scattering extend up to
$\omega \cong |q|$.  Similar behavior is exhibited by  the response functions
of baryons at high density  as they become increasingly relativistic.  In
models HS* and  HS3, the neutrons and protons are moderately relativistic,  due
to  their relatively small effective masses,  which leads to flatter and wider
responses, in contrast to the results for models  GM3 and GM3*,  where
relativistic effects are important only for the leptons.   For
$\omega/\omega_{max} \ll 1$, the response functions are linear with slopes
proportional to $E_{Fi}^*/|q|$.  For non-relativistic particles,  the response
is linear up to  $\omega = \omega_{max}$, with a sharp kinematical cut-off
thereafter.  For relativistic particles, the response  quickly becomes
non-linear with $\omega$ and exhibits a maximum at $\omega$ somewhat less  than
$\omega_{max}$.
\begin{figure}
\begin{center}
\leavevmode
\epsfxsize=3.0in
\epsfysize=3.5in
\epsffile{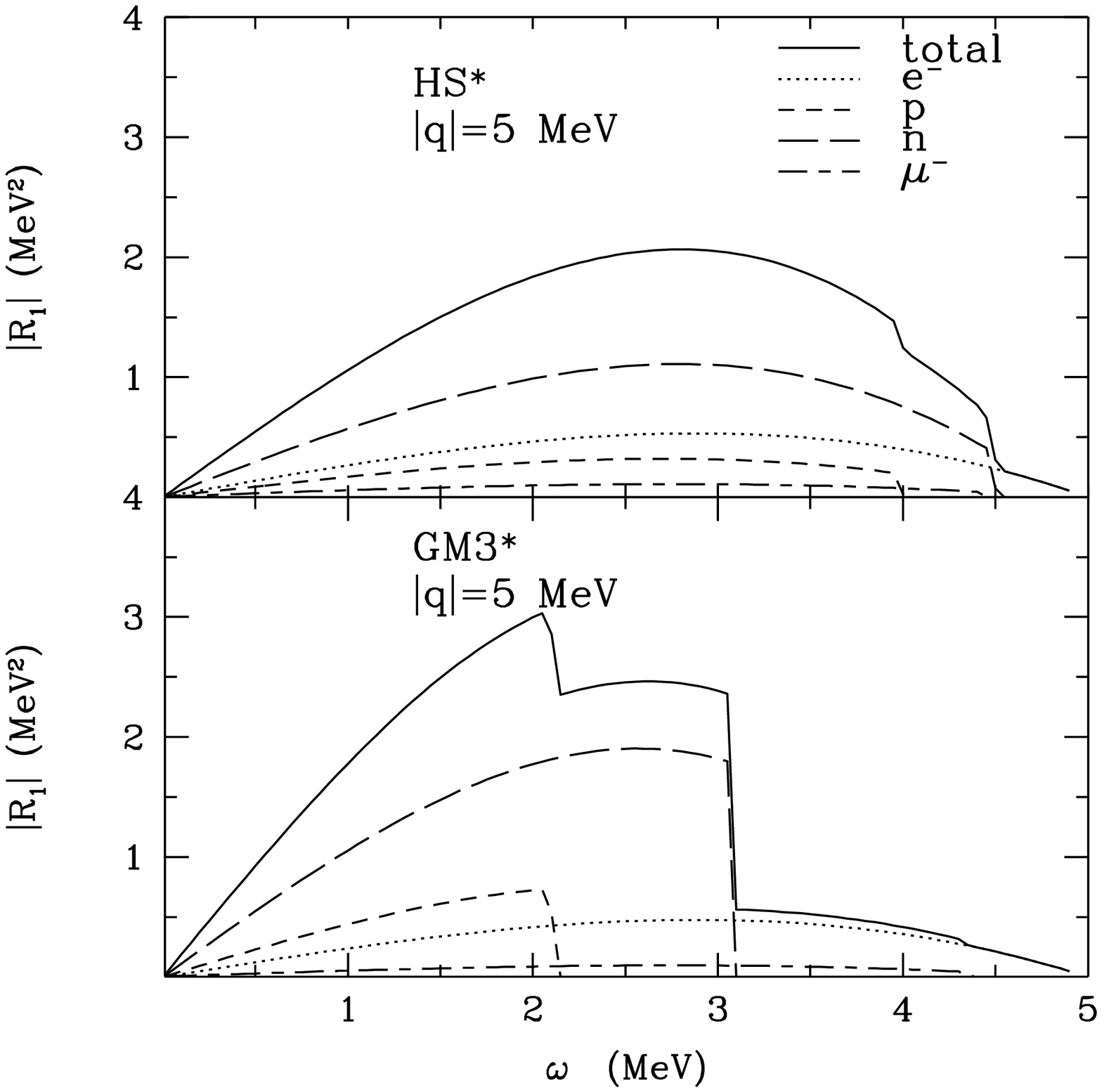}
\leavevmode
\epsfxsize=3.0in
\epsfysize=3.5in
\epsffile{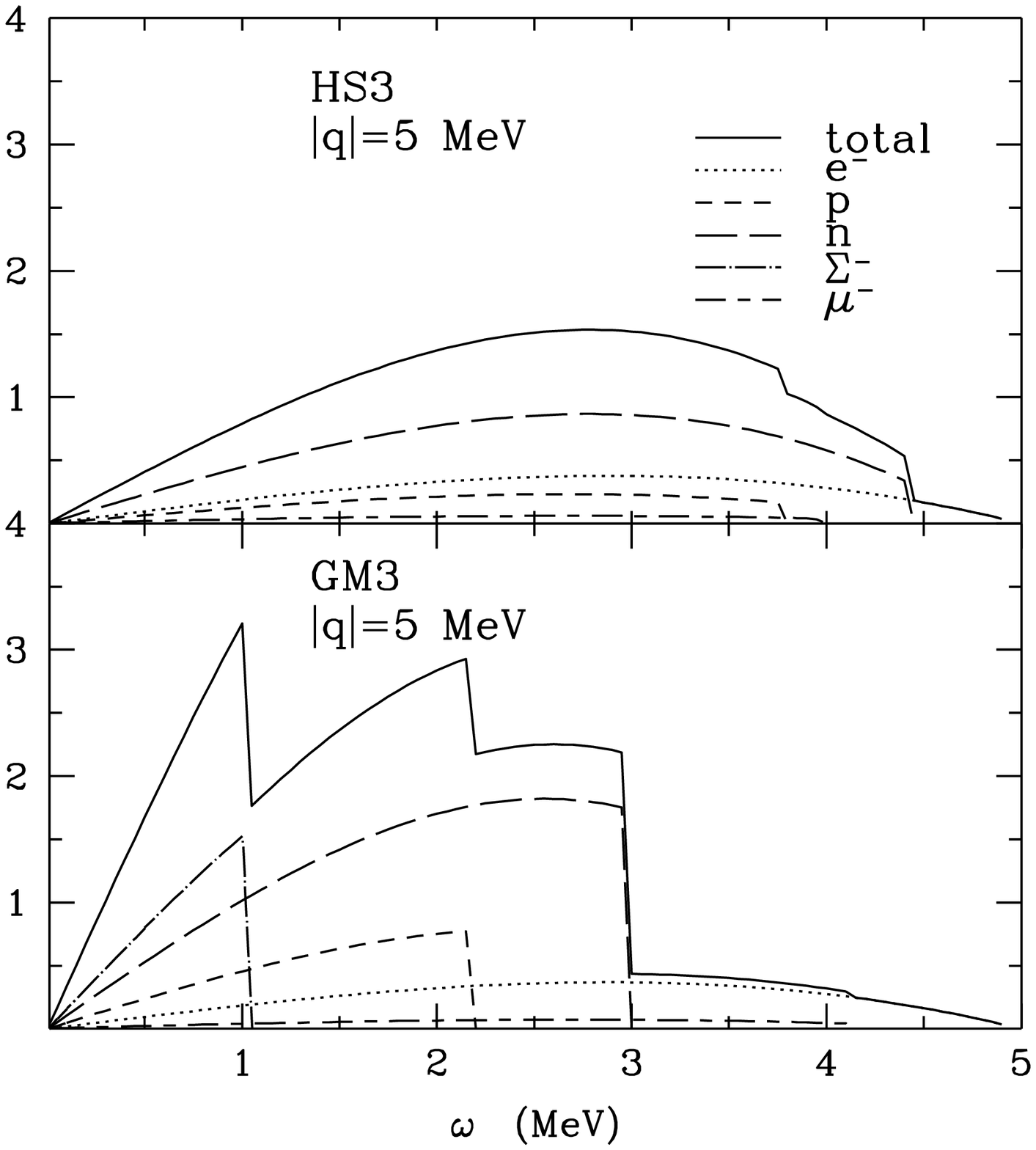}
\caption[]{\footnotesize
-Response function $R_1$ from Eq.~(\ref{r1}),
as a function of the energy transfer $\omega$ for $n_b=0.4~{\rm fm}^{-3}$.
Models HS* and GM3* in the left panels are for matter with nucleons and
leptons only, while models HS3 and GM3 in the right panels include hyperons.}
\end{center}
{\label{figr1}}
\end{figure}

The bulk of the response is provided by the baryons. At $n_b=0.4~{\rm
fm}^{-3}$, model HS3 predicts only the $\Lambda$ hyperon to be present and
that the $n$ and $\Lambda$ abundance are nearly the same.  However, at tree
level, neutrinos do not couple to the $\Lambda$,  and thus the bulk of the
response is from the neutrons.  In contrast, model GM3 allows only the
$\Sigma^-$ hyperon to be present  (up to $n_b=0.4~{\rm fm}^{-3}$), to which
neutrinos couple.  Since  $M_{\Sigma^-}^* > M_n^* $, the $\Sigma^-$ response
is larger than that of the neutron for $\omega$ less than the $\omega_{max}$ of
the $\Sigma^-$.   These results clearly show that the response is sensitive to
both the charge and the abundance of strangeness-bearing components in matter.
Further, the response of matter containing strange baryons  differs (right
panels) significantly from that of matter with nucleons only (left panels).

In Figure~4, we show the differential cross sections per unit
volume  from Eq.~(\ref{dcross}), for $\nu_e$ scattering in the different
models shown in Table~4, and, for  $n_b=0.4~{\rm fm}^{-3}$.  The results,
which
receive contributions from all three response functions in Eqs.~(\ref{r1})
through (\ref{r3}),  highlight the role of the composition and of the effective
mass $M^*$, both of which vary between the different models.   The structure in
these cross sections may be easily understood in terms of the various baryonic
components  present. For example, in models GM1 and HS1, the particles present
are $(\Sigma^-,p,n)$ and $(\Sigma^-,p,\Lambda,n)$, respectively, in increasing
order of $k_{Fi}$.  In contrast, for models GM3 and HS3, we have
$(\Sigma^-,p,n)$ and $(p,\Lambda,n)$, respectively. The $\Sigma^-$, when
present, enhances the cross sections  for $\omega < 1$ MeV.  The signature of
the different effective masses in the various models is evident both from the
magnitudes and the shapes of the differential cross sections.

\begin{figure}
\begin{center}
\leavevmode
\epsfxsize=3.0in
\epsfysize=3.5in
\epsffile{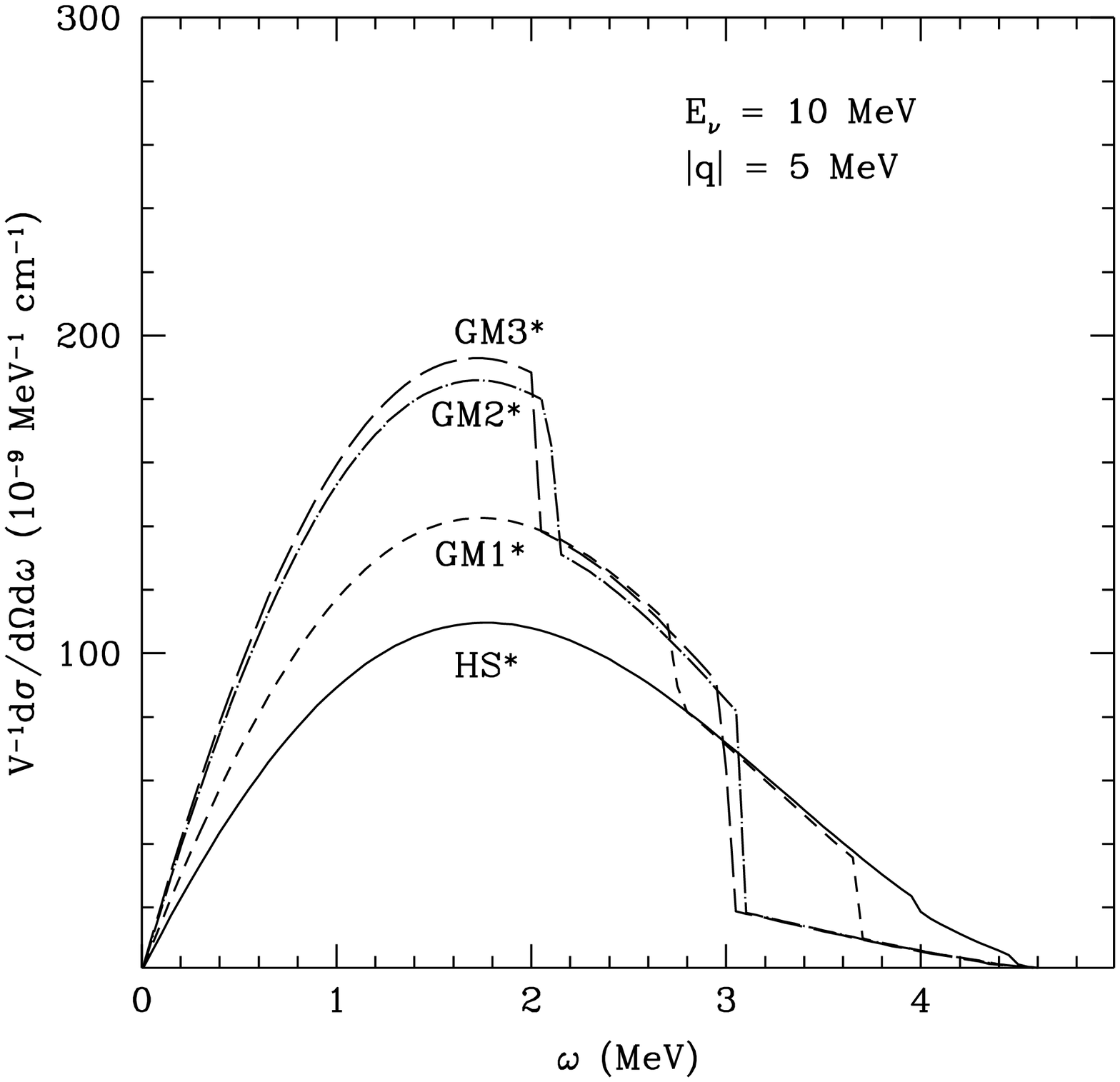}
\leavevmode
\epsfxsize=3.0in
\epsfysize=3.5in
\epsffile{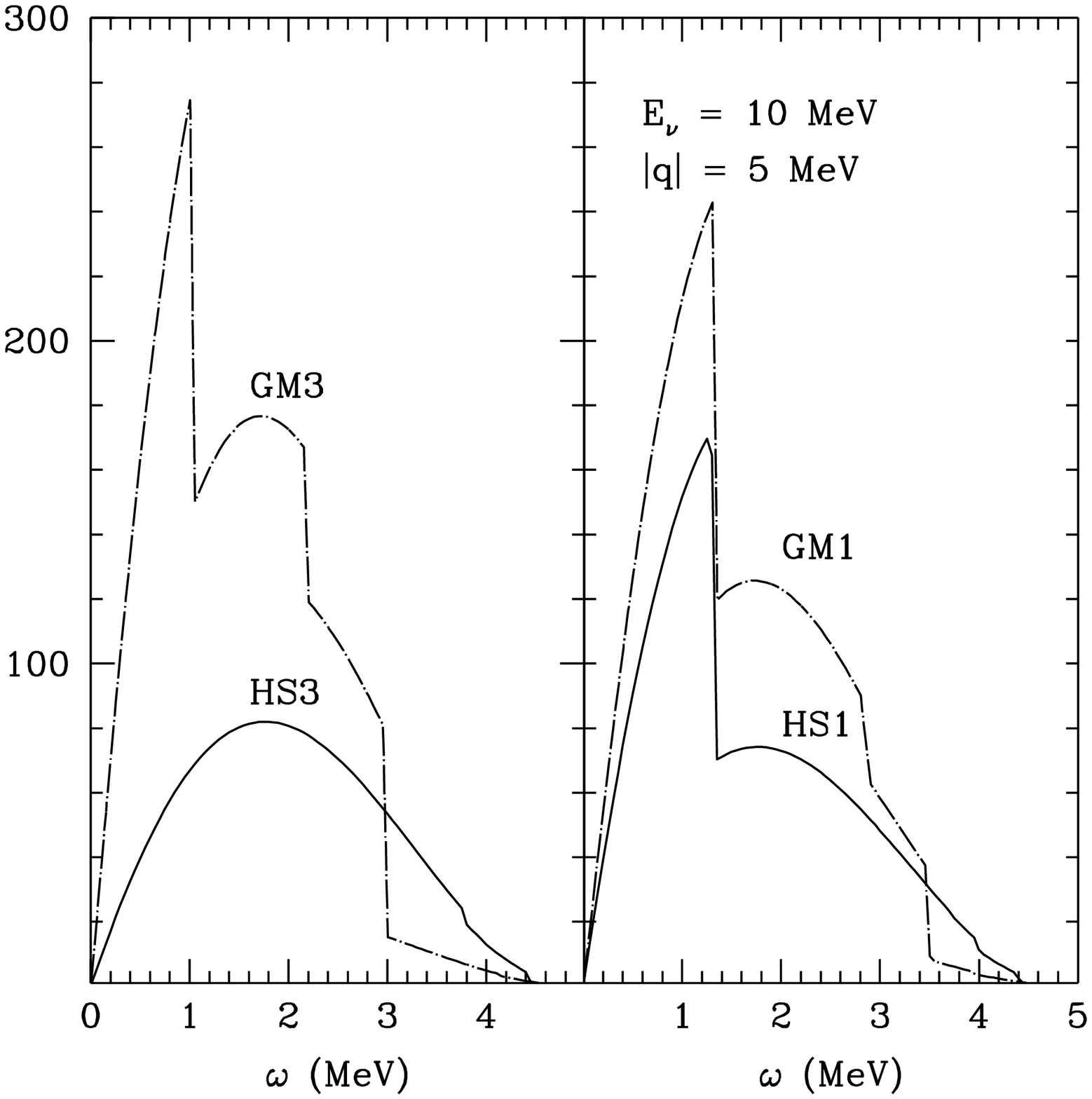}
\end{center}
\caption[]{\footnotesize
Differential cross sections from Eq.~(\ref{dcross}), at $n_b=0.4~{\rm
fm}^{-3}$.  In all models except HS3, the contribution from the $\Sigma^-$
hyperon  is clearly seen in the low energy transfer region. Model designations
are the same as in Figure~3.}
{\label{difcross}}
\end{figure}

A qualitative understanding of the differential cross sections may be obtained
by considering the non-relativistic approximation for the baryon polarization
$\Pi^{\alpha\beta}$. In this limit, only ${\rm Im}~\Pi^{00}$ contributes,
giving
\begin{eqnarray}
\frac{1}{V}\frac{d^3\sigma}{d\Omega^{'2} dE_{\nu}^{'}} =
\frac{G^2}{4\pi^3}~ \frac{E_{\nu}^{'}}{E_{\nu}} |q|^2 ~
A \sum_i [ C_{Vi}^2 S(q,\omega) + 3C_{Ai}^2 {\cal S}(q,\omega) ] \,,
\end{eqnarray}
where $S(q,\omega)$ is the scalar density response function and   ${\cal
S}(q,\omega)$ is the spin density response  function.  This result has been
obtained earlier by Iwamoto \& Pethick (1982).  In the non-relativistic limit,
$S(q,\omega) =  {\cal S}(q,\omega)$ in the mean field models, since there are
no explicit spin-dependent interactions.  For small energy transfers,
\begin{eqnarray*}
S(q,\omega)= \frac{M^{*2}}{2\pi|q|}\omega\,, \qquad {\rm for}~~0\le\omega\le
\frac{k_f}{M^*} |q| \,.
\end{eqnarray*}
Thus, at low density, when $M_i^* \gg k_{Fi}$, the magnitude of the
differential cross sections is essentially determined by the effective mass.
Note, however, that the total cross section does not depend on $M^*$ in this
limit,  since $S(q,\omega)$ depends linearly on $\omega$ and the kinematical
cutoff on $\omega$ is inversely proportional to $M^*$.   This is in accordance
with the sum rule for density-density fluctuations in the non-relativistic
limit.

The total cross section per unit volume is evaluated by integrating
Eq.~(\ref{dcross2}) over the kinematically allowed region.   At zero
temperature, only  $\omega\ge 0$ contributes, due to Pauli blocking and
$\omega
\le|q|\le (2E_{\nu}-\omega)$,  where $ E_{\nu}$ is the incident neutrino
energy.  The density dependence of the cross sections is shown in
Figure~5.  The appearance of the $\Sigma^-$ increases  the
cross section relative to that in matter without the $\Sigma^-$.  However, the
appearance of the $\Lambda$, which furnishes baryon number, results in a
decrease of the  cross section, since the neutron abundance is decreased.  At
high density, when the baryons become increasingly relativistic, the total
cross sections begin to show sensitivity to $M_i^*$, unlike in the
non-relativistic limit,  where such a dependence is absent.  This partly
accounts for the  differences in the cross sections between the various models.
\begin{figure}
\begin{center}
\leavevmode
\epsfxsize=3.0in
\epsfysize=3.5in
\epsffile{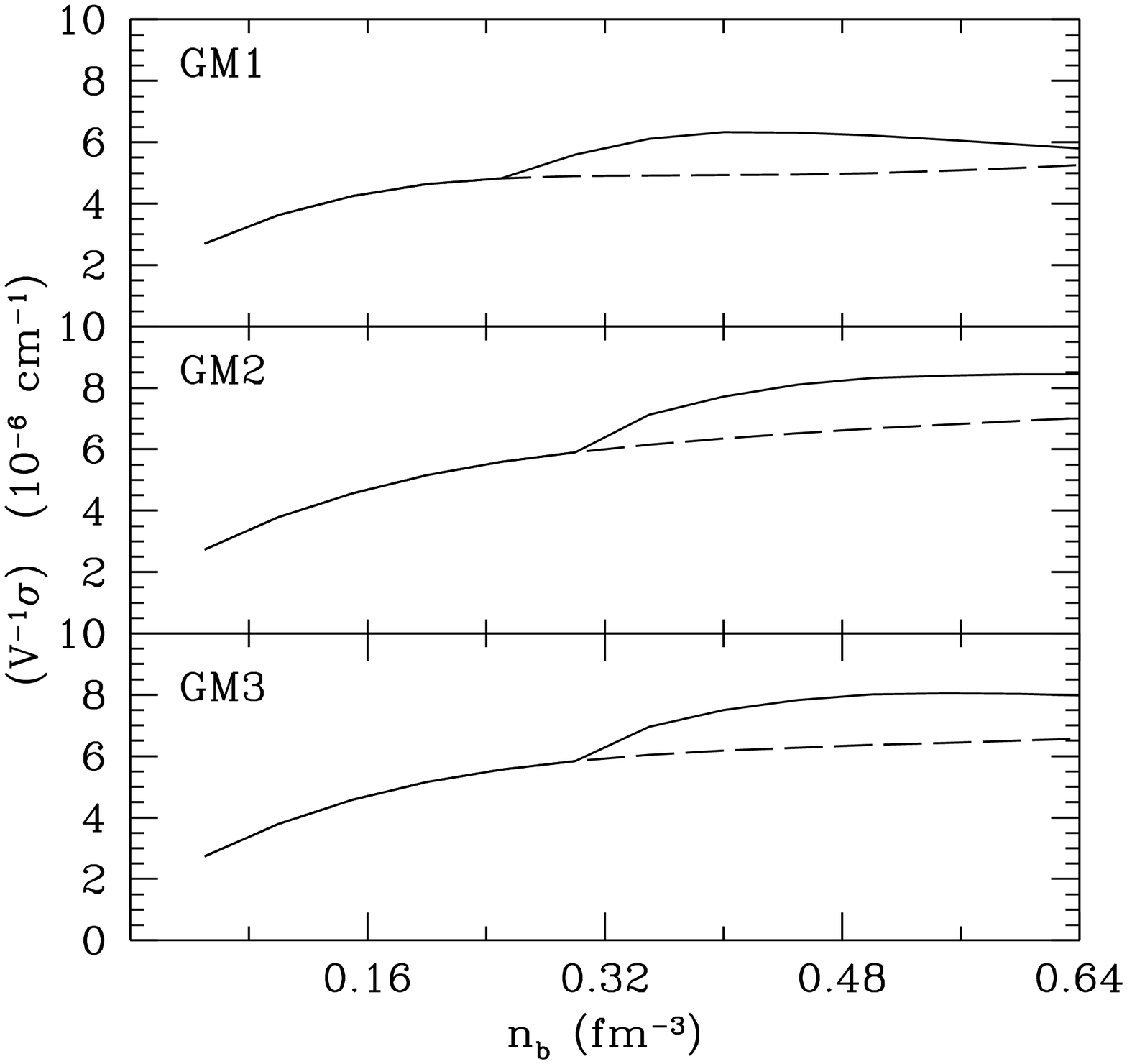}
\leavevmode
\epsfxsize=3.0in
\epsfysize=3.5in
\epsffile{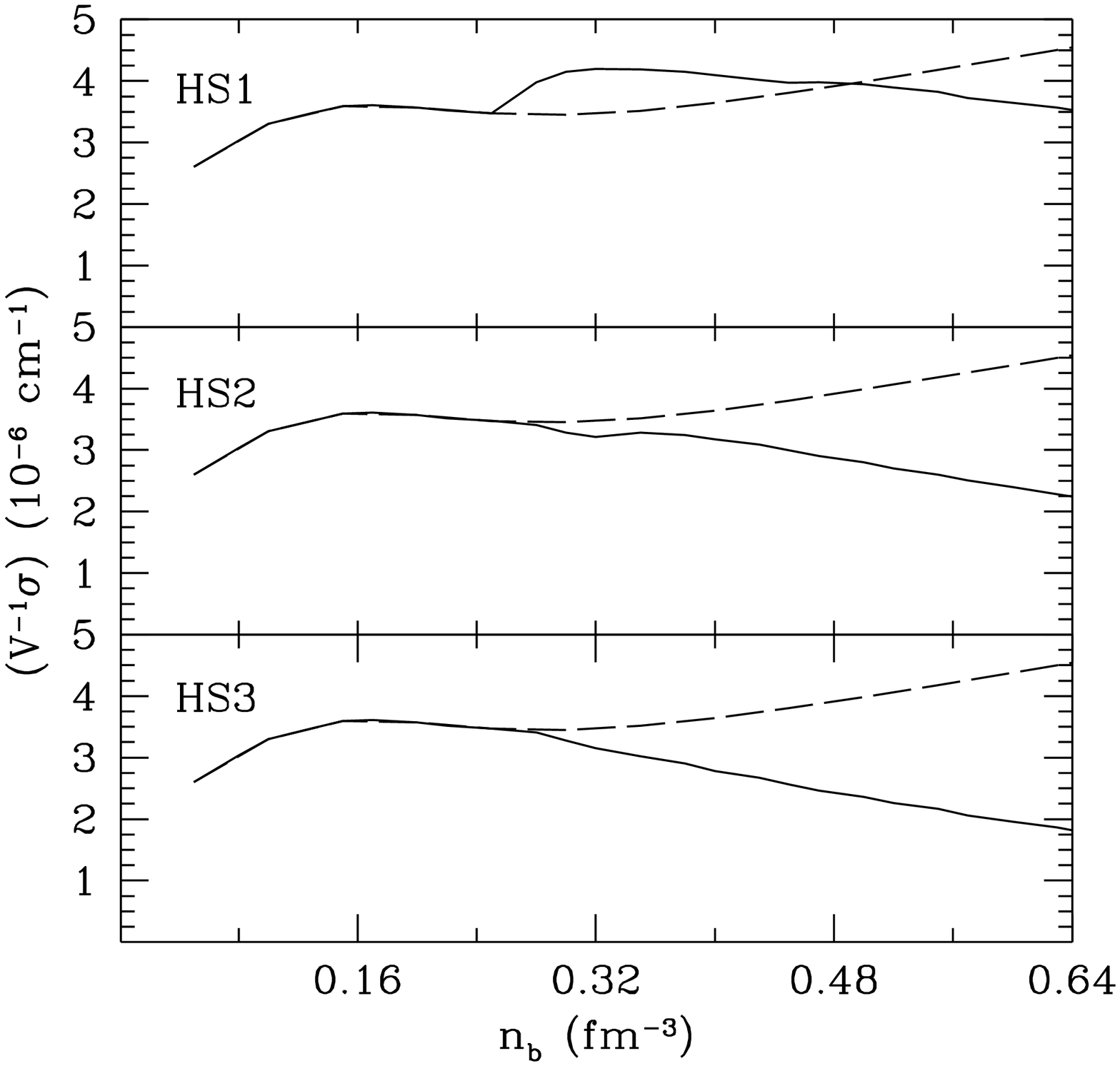}
\end{center}
\caption[]{\footnotesize
The total cross section per unit volume for 10 MeV electron neutrinos
for the models discussed in \S 2.  Solid lines show results with hyperons
and  dashed lines show results with nucleons only.
}
\label{tcross}
\end{figure}

Note also the large suppression (almost a factor of 2 at high density) in
the cross sections for models without scalar self-coupling (HS) compared to
those with such couplings (GM).

The results in Figure~5 give inverse collision mean free paths of 10 MeV
neutrinos interacting via neutral current reactions.  The transport and energy
degradation mean free paths are, in general, different from the collision mean
free path. The former involve appropriate $q$ or $\omega$ weighted  integrals
of the  differential cross sections.  However,  for neutrino diffusion, the
collision  mean  free path is expected to be similar in magnitude to the
transport mean  free paths~(Goodwin \& Pethick 1982).

Figure 6 shows the energy dependence of the cross sections, scaled by
$E_\nu^3$, for different baryon densities.   As expected for zero temperature
matter,  the cross sections in both cases vary essentially as $E_\nu^2\cdot
E_\nu$, where the first factor arises from the basic neutrino cross section on
a single baryon and the second factor arises from the number of participating
particles of a given species.  Note, however, that at high density, the baryon
number density dependence is non-monotonic for the model HS1. This may be
understood in terms of the rapidly decreasing effective masses in this case.

\begin{figure}
\begin{center}
\leavevmode
\epsfxsize=3.0in
\epsfysize=3.5in
\epsffile{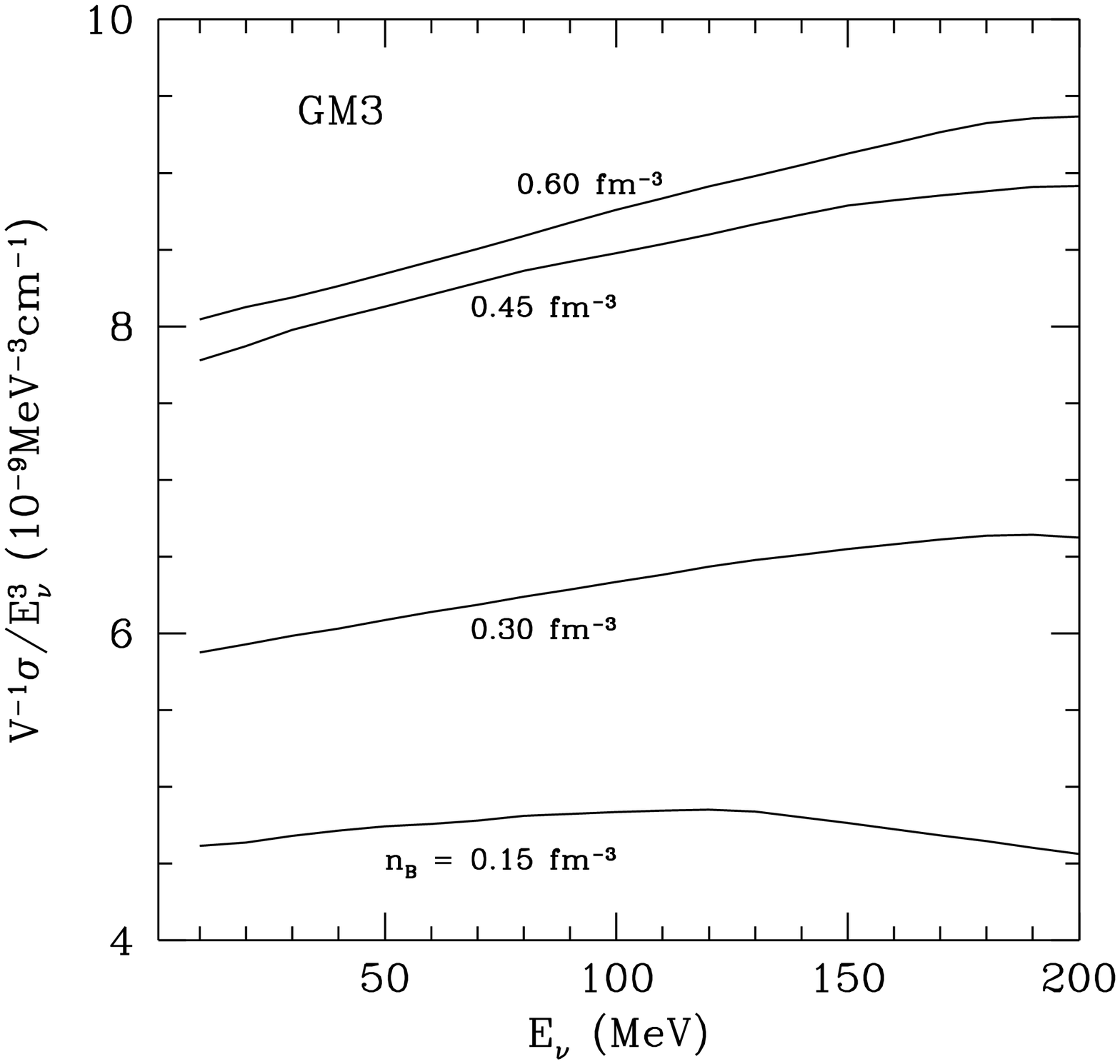}
\leavevmode
\epsfxsize=3.0in
\epsfysize=3.5in
\epsffile{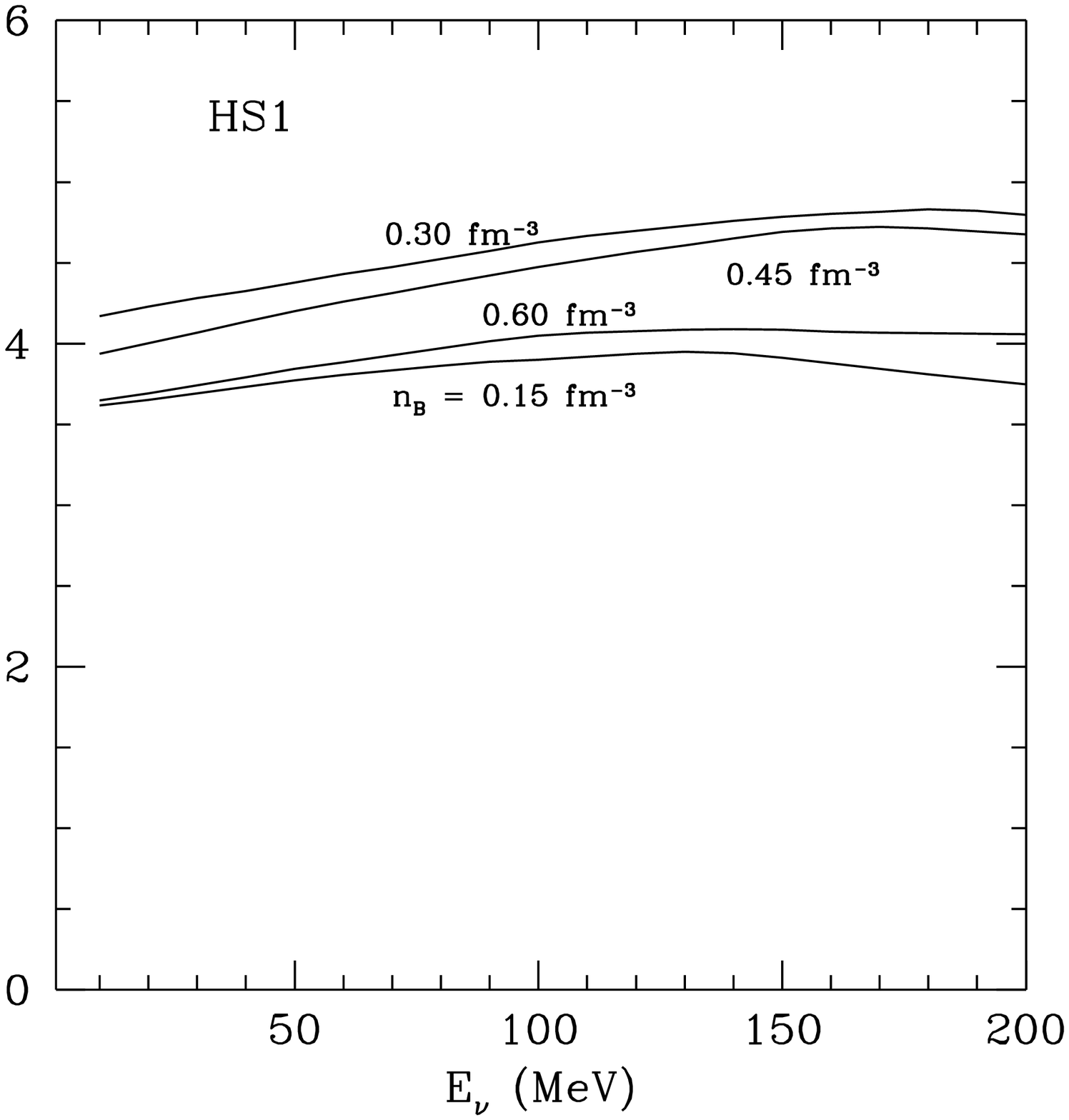}
\end{center}
\caption[]{\footnotesize
Energy dependence of the scaled (with $E_\nu^3$)
total cross section per unit volume as a function of neutrino energy $E_\nu$.
}
\label{ecross}
\end{figure}

It is worthwhile to mention that, in a nuclear medium, the effective value of
$|C_A|$ is quenched (Wilkinson 1973; Rho 1974).  At the nuclear equilibrium
density $n_0=0.16~{\rm fm}^{-3}$, $|C_A| \simeq 1$, and it is expected to
remain at  approximately this value for densities not too far above $n_0$.
Modifications of this nature require that the calculations be carried out
beyond the mean field level for the strongly interacting particles.  A first
orientation of such effects may, however, be obtained by using $C_A=1$ in the
case of nucleons-only matter in the present calculations.  In this case, the
cross sections are reduced by about (15-20)\%.  At present, however,
it is not known if the presence of hyperons at $n_b >> (2-3)n_0$ brings about
additional modifications of $|C_A|$.

\section{ Conclusions }

Many calculations of the composition of dense matter indicate that
strangeness-rich matter should be present in the core of neutron stars.  In
this work, we have identified neutral current neutrino interactions with
hyperons that are important sources of opacity.  We find that significant
contribution to the neutrino opacity arises from scattering involving
$\Sigma^-$ hyperons.  Although the lowest order tree level contributions from
the  neutral $\Lambda$ and $\Sigma^0$ vanish, these particles, when present,
furnish baryon number which decreases the relative concentrations of nucleons.
This leads to significant reductions in opacity.

The neutrino cross section depends sensitively on the Fermi momenta and
effective masses of the various particles present in matter.  Whether or not a
particular hyperon is present depends on the many-body description of charge
neutral beta-equilibrated matter.  We find that as long as one or the other
hyperon is present, the cross sections are significantly modified from the case
of nucleons-only matter.  Strong interactions of strange particles in matter is
currently poorly known, our knowledge being restricted to regions near nuclear
equilibrium density.   In field theoretical descriptions of dense matter,
relatively small changes in the hyperon couplings lead to large differences in
the composition and the effective masses of the various particles.
This emphasizes the need to pin down the mass shifts of baryons in dense
matter, particularly at high density (for initial attempts,
see, for example, Savage \& Wise 1995).  In this work, we have explored the
extent to which the cross sections are modified due to the prevalent
uncertainty in strong interactions.   Depending on the behavior of the
effective masses at high density, the opacities may vary by a factor of about 2
at high density.

Our findings here suggest several directions for further study.  The extension
of these calculations to finite temperature and to include correlations between
the different particles is straightforward and is under progress.   During its
early evolution, the protoneutron star attains entropies per baryon of order
1-2 and central temperatures of order 30-50 MeV. Such matter is  degenerate,
since the chemical potentials of the constituents are typically of order
200-300  MeV (Prakash et al. 1995).  Hence, the zero temperature results are
not
expected to be  greatly modified.  The presence of charged  particles, such as
the $\Sigma^-$, could make available low energy collective plasma modes through
electromagnetic correlations, in addition to the scalar, vector and iso-vector
correlations.  Effects of strangeness on lepton number and energy transport may
be studied by employing energy averages (Rosseland means) of the opacities  in
present protoneutron codes.  Tables of such useful average opacities will be
made generally available.  With new generation neutrino detectors capable of
recording thousands of  neutrino events, it may be possible to distinguish
between different scenarios observationally.

\vspace*{0.3in}

This work was supported in partly the U.S. Department of Energy under
contract number DOE/DE-FG02-88ER-40388 and by the NASA grant NAG 52863.  We
thank Jim Lattimer and Chuck Horowitz for helpful discussions.

\bigskip
{\Large \centerline {\bf {References} } }
\begin{tabbing}
xxxxx\=xxxxxxxxxxxxxxxxxxxxxxxxxxxxxxxxxxxxxxxxxxxxxxxxxxxxxxxxxxxxxxxxxxxxxxxx\kill
\ni Bionta, R. M., et al., 1987, Phys. Rev. Lett. 58, 1494 \\
\ni Bruenn, S., 1985, ApJ. Suppl. 58, 771 \\
\ni Burrows, A., 1988, ApJ 334, 891 \\
\ni -------, 1990, Ann. Rev. Nucl. Sci. 40, 181 \\
\ni Burrows, A \& Lattimer, J. M., 1986, ApJ 307, 178 \\
\ni Burrows, A \& Mazurek, T. J., 1982, ApJ 259, 330 \\
\ni Cooperstein, J., 1988, Phys. Rep. 163, 95 \\
\ni Ellis, J., Kapusta, J. I., \& Olive, K. A. 1991,
Nucl. Phys. B348, 345 \\
\ni Ellis, P. J., Knorren, R., \& Prakash, M., 1995, Phys. Lett. B349, 11 \\
\ni Gaillard, J.-M., \& Sauvage, G. 1984, Ann. Rev. Nucl. Part. Sci., 34, 351
\\
\ni Glendenning, N. K., \& Moszkowski, S. A. 1991,
Phys. Rev. Lett. 67, 2414 \\
\ni Goodwin, B. T., 1982, ApJ 261, 321 \\
\ni Goodwin, B. T. \& Pethick, C. J. 1982, ApJ 253, 816 \\
\ni Hirata, K., et al., 1987, Phys. Rev. Lett. 58, 1490 \\
\ni Horowitz, C. J., 1992, Phys. Rev. Lett. 69, 2627 \\
\ni Horowitz, C. J., \& Serot, H. S. 1981, Nucl. Phys. A368, 503 \\
\ni Horowitz, C. J., \& Wehrberger, K., 1991a, Nucl. Phys. A531, 665 \\
\ni --------, 1991b, Phys. Rev. Lett. 66, 272 \\
\ni --------, 1992, Phys. Lett. B226, 236 \\
\ni Iwamoto, N. 1982, Ann. Phys. 141, 1                           \\
\ni Iwamoto, N., \& Pethick, C. J., 1982, Phys. Rev. D25, 313 \\
\ni Kaplan, D. B., \& Nelson, A. E. 1986, Phys. Lett. B175, 57      \\
\ni ------, 1986, Phys. Lett. B179, 409 (E)                            \\
\ni Kapusta, J. A., \& Olive, K. A., 1990, Phys. Rev. Lett. 64, 13      \\
\ni Keil, W., \& Janka, H. T., 1995, Astron. \& Astrophys. 296, 145 \\
\ni Knorren, R., Prakash, M., \& Ellis, P. J., 1995, Phys. Rev. C52, 3470 \\
\ni Lamb, D. Q., 1978, Phys. Rev. Lett. 41, 1623 \\
\ni Lamb, D. Q., \& Pethick, C. J., 1976, ApJ 209, L77 \\
\ni Mare\v{s}, J., Friedman, E., Gal. A., \& Jennings, B. K., 1995, preprint
nucl-th/9505003 \\
\ni Maxwell, O., 1987, ApJ 316, 691 \\
\ni Mazurek, T. J., 1975, Astrophys. Space. Sci. 35, 117 \\
\ni Prakash, M., Prakash, Manju, Lattimer, J. M., \& Pethick, C. J.
1992, ApJ 390, L77 \\
\ni Prakash, M., Cooke, J., \& Lattimer, J. M., 1995, Phys. Rev. D52, 661 \\
\ni Prakash, M., Bombaci, I., Prakash, Manju, Ellis, P. J., \\
\> Lattimer, \& J. M., \& Knorren, R., 1995, Phys. Rep., To be submitted \\
\ni Reddy, S., and Prakash, M., 1995, Proc. of the 11th Winter Workshop on
Nuclear \\ \> Dynamics, Key West, Fl, Feb 11--18, 1995 \\
\ni Rho., M. 1974, Nucl. Phys, A231, 493.
\ni Sato, K., 1975, Prog. Theor. Phys. 53, 595 \\
\ni Savage, M., \& Wise, M., 1995, Phys. Rev. D, in press \\
\ni Sawyer, R. F., 1975, Phys. Rev. D11, 2740 \\
\ni ------, 1989, Phys. Rev. C40, 865 \\
\ni ------, 1995, Phys. Rev. Lett. C75, 2260 \\
\ni Sawyer, R. F. \& Soni, R., 1979, ApJ 230, 859 \\
\> ed. J. W. Negele \& E. Vogt (New York: Plenum) \\
\ni Serot, B. D., \& Walecka, J. D., 1986, in Adv. Nucl. Phys. 16, \\
\ni Thorsson, V., Prakash, M., \& Lattimer, J. M. 1994, Nucl. Phys. A572, 693
\\
\ni Tubbs, D. L., \& Schramm, D. N., 1975, ApJ 201, 467 \\
\ni van den Horn. L. J., \& Cooperstein, J., 1986, ApJ 300, 142 \\
\ni Wilkinson, D. H., 1973, Phys. Rev. C7, 930. \\
\end{tabbing}

\end{document}